\documentclass[apj]{emulateapj}

\usepackage{times}

\shorttitle{Line Widths in Solar Active Regions}
\shortauthors{Brooks \& Warren}

\begin{document}


\title{Measurements of Non-Thermal Line Widths in Solar Active Regions}

\author{David H. Brooks\altaffilmark{1,3} and Harry P. Warren\altaffilmark{2}}

\affiliation{\altaffilmark{1}College of Science, George Mason University, 4400 University Drive,
  Fairfax, VA 22030 USA}
\affiliation{\altaffilmark{2}Space Science Division, Naval Research Laboratory, Washington, DC
  20375 USA}
\altaffiltext{3}{Present address: Hinode Team, ISAS/JAXA, 3-1-1 Yoshinodai, Chuo-ku, Sagamihara, 
Kanagawa 252-5210, Japan}


\begin{abstract}
 Spectral line widths are often observed to be larger than can be accounted for by thermal and
 instrumental broadening alone. This excess broadening is a key observational constraint for both
 nanoflare and wave dissipation models of coronal heating. Here we present a survey of non-thermal
 velocities measured in the high temperature loops (1--4\,MK) often found in the cores of solar
 active regions. This survey of \textit{Hinode} Extreme Ultraviolet Imaging Spectrometer (EIS)
 observations covers 15 non-flaring active regions that span a wide range of solar conditions. We
 find relatively small non-thermal velocities, with a mean value of 17\,km~s$^{-1}$, and no
 significant trend with temperature or active region magnetic flux. These measurements appear to
 be inconsistent with those expected from reconnection jets in the corona, chromospheric
 evaporation induced by coronal nanoflares, and Alfv\'{e}n wave turbulence models. Furthermore,
 because the observed non-thermal widths are generally small, such measurements are difficult and
 susceptible to systematic effects.
\end{abstract}



\section{introduction}

The solution to the problem of how the solar corona maintains its high temperature relative to the
cool solar photosphere is still unresolved. There have been many explanations put forward, and
there are several reviews on the topic in the literature from varying perspectives
\citep[e.g.][]{mandrini_etal2000,walsh_2003,klimchuk_2006,reale_2010,parnell_2012,arregui_2015}.
Two of the most studied ideas are based on magnetic reconnection and MHD waves. In the
reconnection scenario, the magnetic field in the corona is braided by turbulent convection in the
photosphere, and the energy is released through small-scale events popularly known as nanoflares
\citep{parker_1983,parker_1988}. In the MHD wave scenario, the interaction between photospheric
convection and the magnetic field can produce waves that propagate upward along the magnetic field
and dissipate energy through a variety of possible mechanisms such as phase mixing
\citep{heyvaerts&priest_1983}, or resonant absorption \citep{ionson_1978}. More recently,
chromospheric jets have been suggested as a mechanism for supplying mass and energy to the corona
directly from the lower atmospheric layers \citep{depontieu_etal2009,depontieu_etal2011}.

There has been tentative observational evidence for several of the predicted details of these
models.  \citet{depontieu_etal2011}, for instance, show examples of chromospheric spicules
producing an apparent response at transition region and coronal temperatures.
\citet{cirtain_etal2013}, show cases of large scale wrapped and twisted coronal structures that
could be evidence of magnetic braiding.  \citet{testa_etal2013} find evidence of rapid variability
at the footpoints of hot loops that they interpret as signatures of heating associated with
reconnection events occuring in the overlying loops.  Very recently, \citet{okamoto_etal2015}
observed decreasing amplitude wave-like motions in chromospheric images that they suggest are
evidence of resonant absorption. Future observations and comparisons with numerical models will
clarify if these interpretations are correct.

Despite these recent advances, and we have highlighted only a few examples, in general it has been
difficult to discriminate between the theoretical models and reconcile them with all aspects of
the observations. One diagnostic that has the potential to be a good discriminator is the
measurement of non-thermal broadening in excess of the thermal and instrumental broadening of EUV
spectral lines.  As pointed out by \citet{cargill_1996}, the nanoflare-heated corona model
predicts the existence of many reconnection jets in multiple directions along an observed
line-of-sight, and this should lead to significant non-thermal velocities on the order of 250\,km
s$^{-1}$ or larger. In contrast, hydrodynamic models of chromospheric evaporation in response to
coronal nanoflares suggest non-thermal velocities of 20--36\,km s$^{-1}$, increasing with
temperature, at coronal loop tops in the 1.1--5.6\,MK temperature range
\citep{patsourakos&klimchuk_2006}. Models of shock heating driven by Alfv\'{e}n waves also predict
high velocities, $>$ 100\,km s$^{-1}$, see e.g. \citet{antolin_etal2008}, while models of
Alfv\'{e}n wave turbulence \citep{vanballegooijen_etal2011} show non-thermal velocities of
25--35\,km s$^{-1}$ at the tops of loops formed near 1.6\,MK
\citep{asgaritarghi_etal2014}. Furthermore, models that attempt to explain the first ionization
potential (FIP) effect based on the forces acting on propagating waves suggest velocities on the
order of 50--80\,km s$^{-1}$ \citep{laming_2004,laming_2012}.

These are significant differences that are, in principal, observationally detectable, so that at
least some of the possible explanations for the observed non-thermal broadening could be ruled
out, even if the heating mechanism itself cannot be verified.  For example, measurements of
non-thermal velocities of 40\,km s$^{-1}$ or so would tend to support a chromospheric evaporation
or Alfv\'{e}n wave turbulence explanation, though clearly they could not discriminate between
them.  Conversely, the nanoflare model may predict high non-thermal velocities, but these may only
be apparent immediately after the energy release, and could be difficult to detect due to the low
emission measure of any hot plasma, a problem that may be further exacerbated by non-equilibrium
conditions \citep{bradshaw&cargill_2006}.

The presence of a hot active region (AR) plasma component is in fact a central prediction of
impulsive heating models, and significant effort has been expended assessing whether such plasma
can be observed.  See e.g. \citet{reale_etal2009a}, \citet{reale_etal2009b}, and
\citet{schmelz_etal2009} for initial reports of detections, \citet{reale_2014} for a review of
subsequent efforts, and \citet{odwyer_etal2011} and \citet{winebarger_etal2012} for reports of
negative results and the difficulties of observing the appropriate temperature range.
Notwithstanding these problems, measurements of non-thermal velocities at high temperatures (and
any trend with temperature) are clearly significant for establishing whether the nanoflare-heated
corona model is viable.

Previous measurements, however, have indicated that the excess width of EUV and soft X-ray lines
is smaller than suggested by \citet{cargill_1996}. Early observations from Skylab indicated values
of 10--25\,km s$^{-1}$ for UV coronal lines formed near 1.7\,MK such as \ion{Fe}{11} and
\ion{Fe}{12}, for example, see e.g. \citet{cheng_etal1979}. Later measurements of the \ion{Mg}{11}
resonance line formed around 2.8\,MK from the Solar Maximum Mission were found to be somewhat
larger: 40--60\,km s$^{-1}$ \citep{acton_etal1981,saba&strong_1991}, but these are also
significantly smaller than expected from the high temperature reconnection jets. Recent studies
using \textit{Hinode}/EIS \citep[EUV Imaging Spectrometer,][]{culhane_etal2007} and SOHO/SUMER
\citep[Solar Ultraviolet Measurements of Emitted Radiation,][]{wilhelm_etal1995} have succeeded in
observing hot plasma emission from the \ion{Ca}{17} 192.858\,\AA\, and \ion{Fe}{18} 974.86\,\AA\,
lines formed near 5.6\,MK and 7.1\,MK, respectively \citep{ko_etal2009,teriaca_etal2012}.  They did
not, however, assess the non-thermal broadening, and it is difficult for solar observations to
access much higher temperatures outside of flares.

Observations of stellar atmospheres, however, have routinely been made in the 7--11\,MK range with
high spectral resolution instruments on space missions such as the Far Ultraviolet Spectroscopic
Explorer (FUSE) and the Hubble Space Telescope. \citet{redfield_etal2003} have studied the
profiles of the \ion{Fe}{18} 974\,\AA\, and \ion{Fe}{19} 1119\,\AA\, lines, formed around 7.1\,MK
and 8.9\,MK, for a sample of late type stars observed by FUSE.  \citet{linsky_etal1998} used the
HST Goddard High Resolution Spectrograph to examine the profile of the \ion{Fe}{21} 1354\,\AA\,
line, formed near 11.2\,MK, in the atmosphere of the Capella binary. Both of these studies
basically found no appreciable non-thermal line broadening and concluded that all these lines were
consistent with thermal broadening.

Stellar observations of course have no spatial resolution, and that is the advantage of
spectrometers such as EIS that have allowed the measurement of non-thermal velocities in discrete
solar coronal structures. EIS observations have shown that the largest non-thermal velocities in
an AR are located in the regions of lowest intensity
\citep{delzanna_2008,doschek_etal2008,harra_etal2008,baker_etal2009}, where the plasma is
outflowing and may contribute to the solar wind
\citep{sakao_etal2007,harra_etal2008,doschek_etal2008,brooks_etal2015}.  Measurements of the
non-thermal velocities in these regions are about 30--60\,km s$^{-1}$ using lines formed at
1.6--2.2\,MK \citep{delzanna_2008,doschek_etal2008,brooks&warren_2011}, and these values are
somewhat larger than observed in the AR core. \citet{warren_etal2008a} and
\citet{brooks&warren_2009}, for example, obtained values of 20--30\,km s$^{-1}$ in the moss at the
footpoints of 1.6\,MK loops.

Despite these studies, there are relatively few accurate measurements of non-thermal velocities in
AR core loops.  One of the few cases is that of \citet{imada_etal2009}, who performed a very
interesting study of an AR core at the limb, and reported an average value of 13\,km s$^{-1}$ at
2.8\,MK. They studied the AR core as a whole, however, and did not isolate the emission from the
high temperature loops, or investigate any trend with temperature in detail.  Since the region was
also observed at the limb, the mass motions may be diminished due to the line-of-sight. This
effect was noted by \citet{hara_etal2008a}, who found a decrease in the non-thermal velocities at
the footpoints of $\sim$2\,MK loops as the AR they studied tracked from disk center to limb, a
trend that has been suggested to be indicative of Alfv\'{e}n wave heating
\citep{mcclements_etal1991,cargill_1996}.  It would also be interesting to know if the AR studied
by \citet{imada_etal2009} is typical or not.

Here we undertake a systematic survey of the non-thermal velocities in the hot plasma in AR cores,
to assess whether the observations and coronal heating models can be reconciled. We isolate 16
loops in 15 different ARs covering a wide range of solar conditions, and measure the non-thermal
velocities over a range of temperatures from 1.1--3.6\,MK. We consider the magnitudes of the
measurements, and any trend with temperature, both of which are important for nanoflare heating
models.

\begin{figure*}[t!]
\centerline{\includegraphics[width=0.9\textwidth]{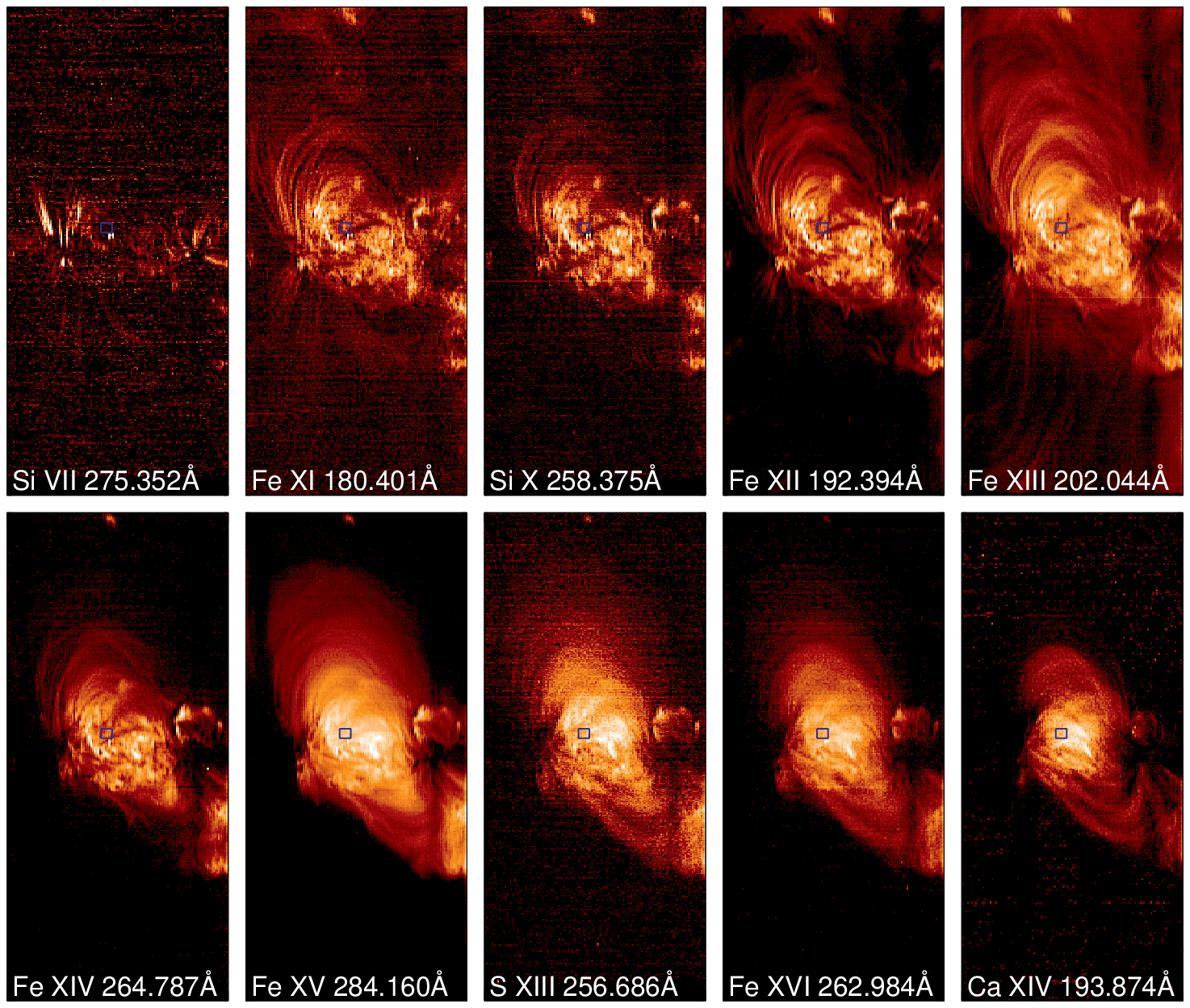}}
\hspace{-0.2in}\centerline{\includegraphics[viewport= 0 540 595
    780,clip,width=1.3\textwidth]{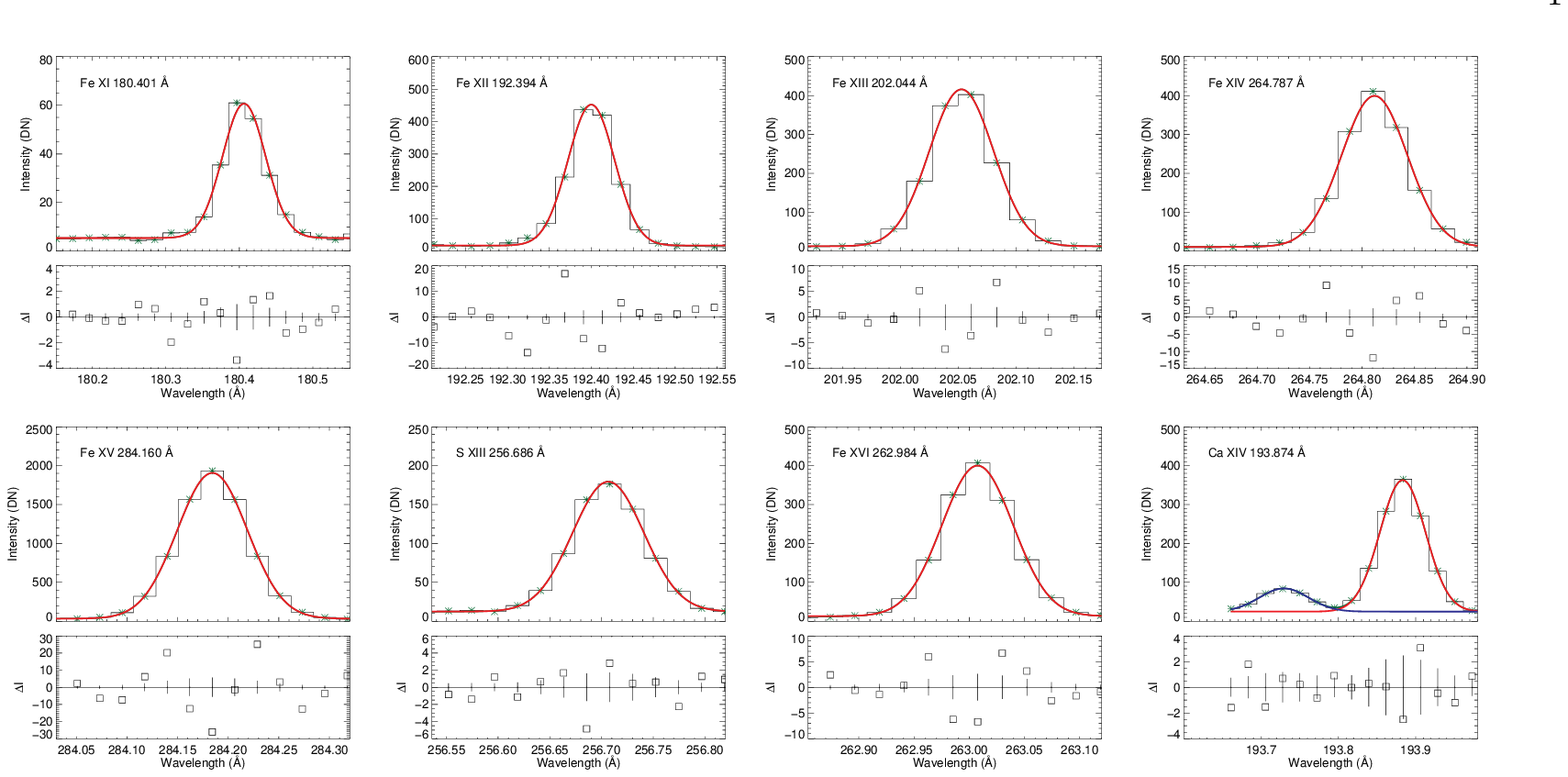}}
\caption{Example spectral images of AR 1193 observed on 2011, April 19. We measured the line
  widths in the high temperature core loops for the region indicated by the blue box. Example fits
  for several of the spectral lines are shown in the lower panels.  }
\label{fig:fig1}
\end{figure*}

\begin{figure*}[t!]
\centerline{\includegraphics[width=0.9\textwidth]{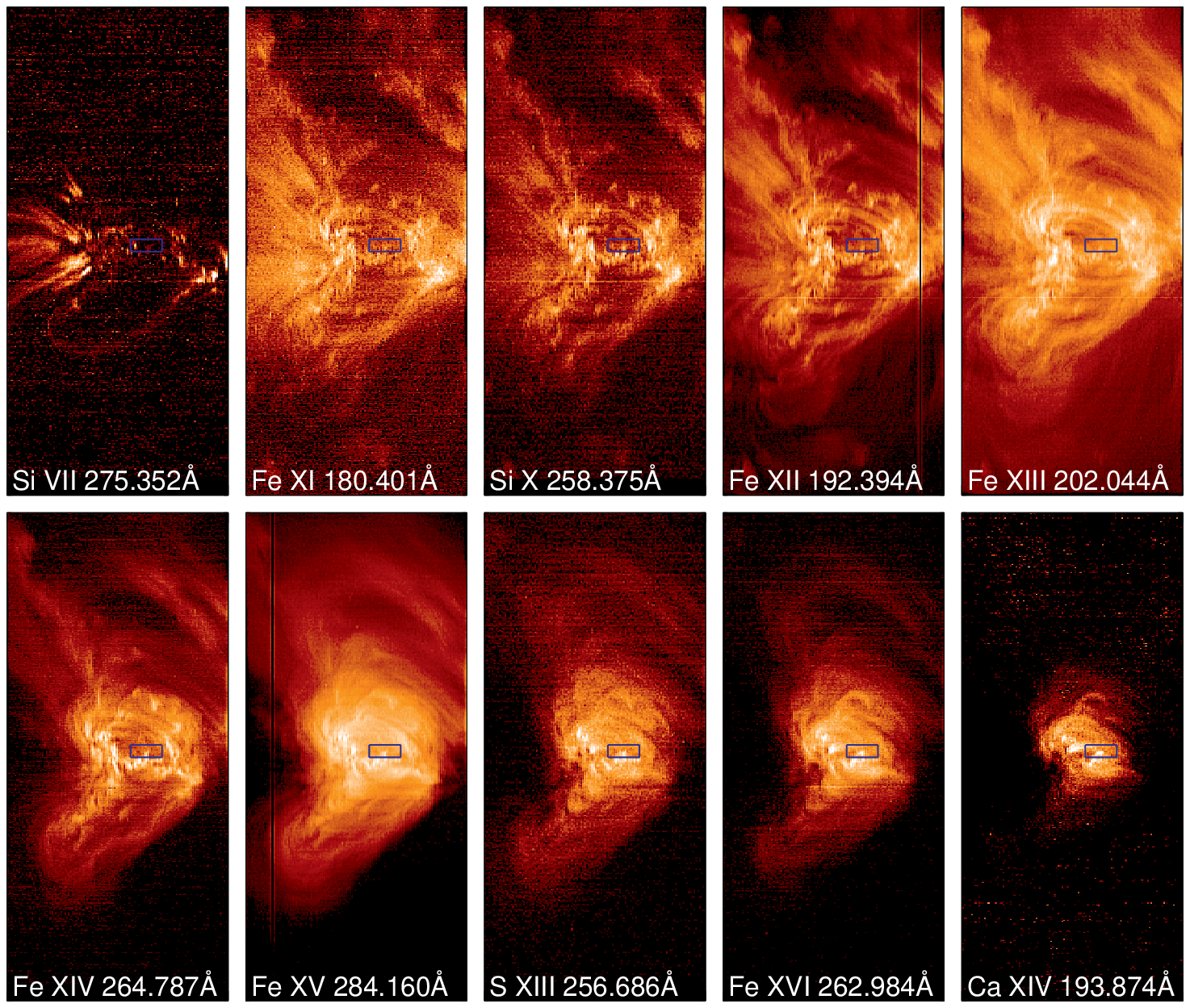}}
\hspace{-0.2in}\centerline{\includegraphics[viewport= 0 540 595
    780,clip,width=1.3\textwidth]{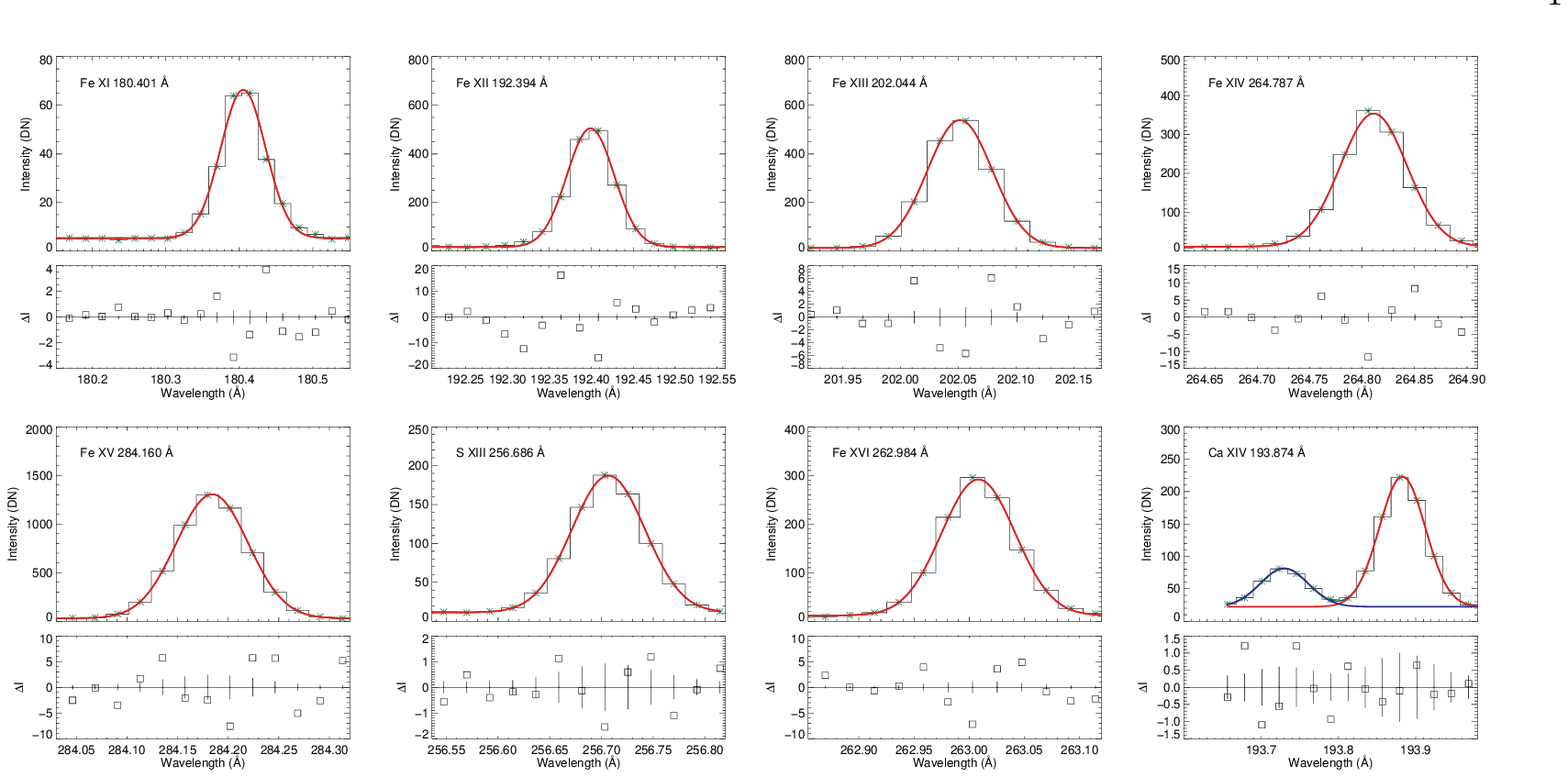}}
\caption{Example spectral images of AR 1190 observed on 2011, April 11. We measured the line
widths in the high temperature core loops for the region indicated by the blue box. Example fits for
several of the spectral lines are shown in the lower panels.
  }
\label{fig:fig2}
\end{figure*}

\section{observations}

\begin{figure*}[t!]
\centerline{\includegraphics[width=\textwidth]{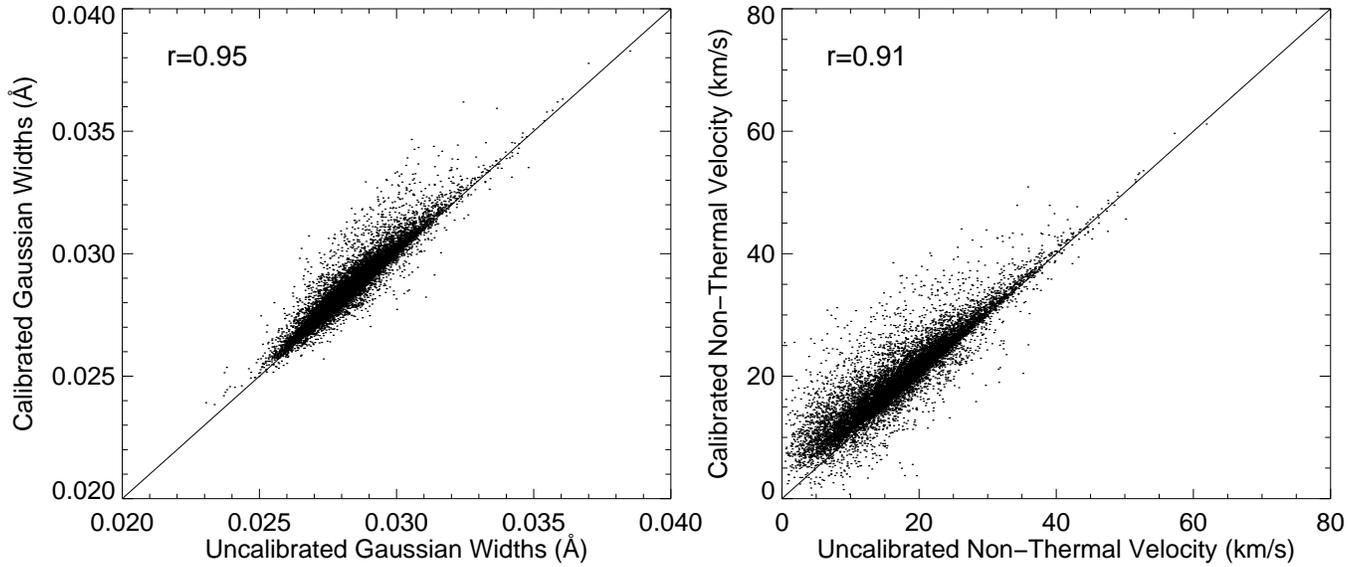}}
\caption{A comparison of line widths derived from Gaussian fits to calibrated and uncalibrated
  line profiles. Gaussian widths are shown on the left and the resulting non-thermal velocities
  are shown on the right. These calculations are for \ion{Fe}{12} 192.394\,\AA\, from the 2011,
  April 19 dataset in Figure \ref{fig:fig1}. Only pixels that are brighter than 10\% of the
  maximum intensity pixel are shown.}
\label{fig:fig3}
\end{figure*}

\begin{figure*}[t!]
\centerline{\includegraphics[angle=90,clip,width=\textwidth]{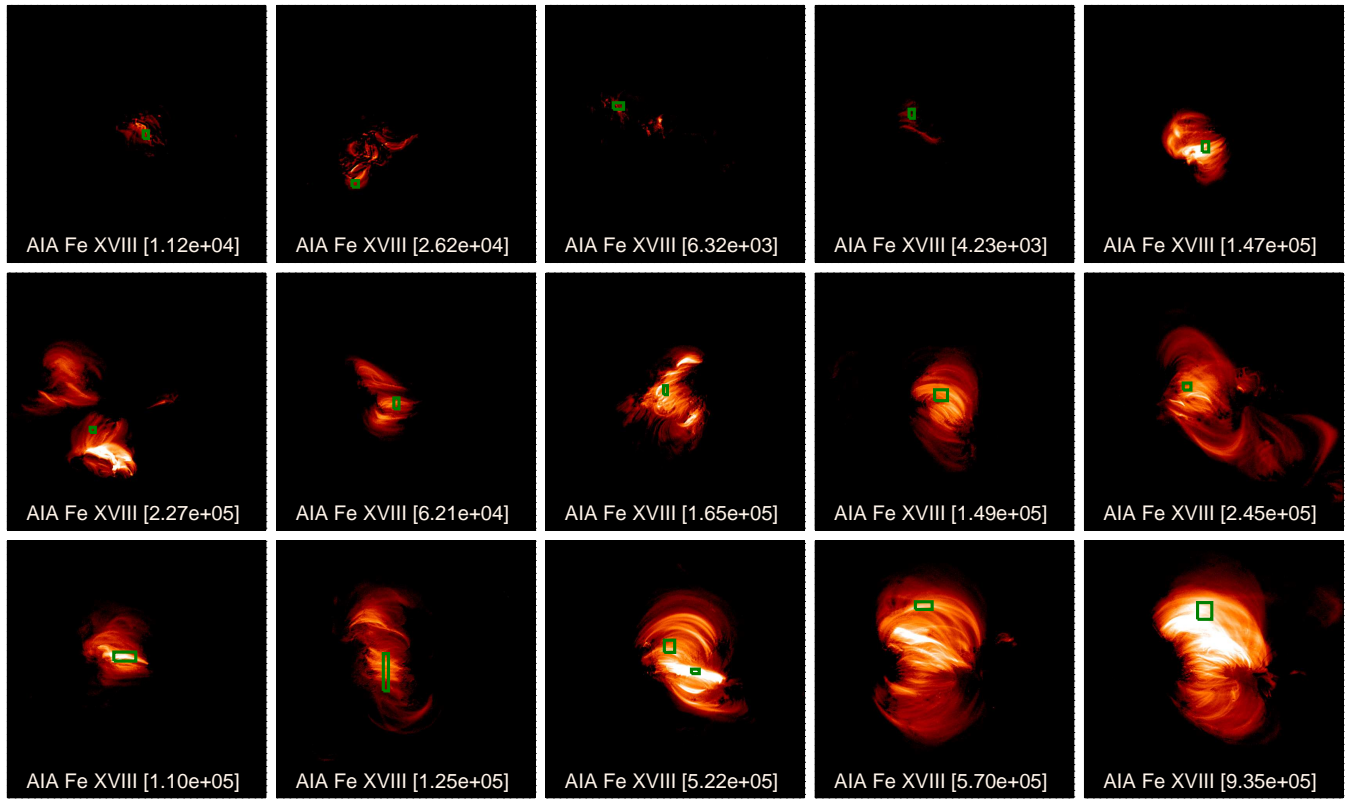}}
\caption{\ion{Fe}{18} 93.92\,\AA\, images of the hot core loops analyzed. These were extracted from the AIA 94\,\AA\, images. 
  The green boxes show the locations where the EIS measurements are made. The number shown is the total intensity in the field-of-view. }
\label{fig:fig4}
\end{figure*}

\begin{figure*}[t!]
\centerline{\includegraphics[viewport = 40 215 585 790,clip,width=1.00\textwidth]{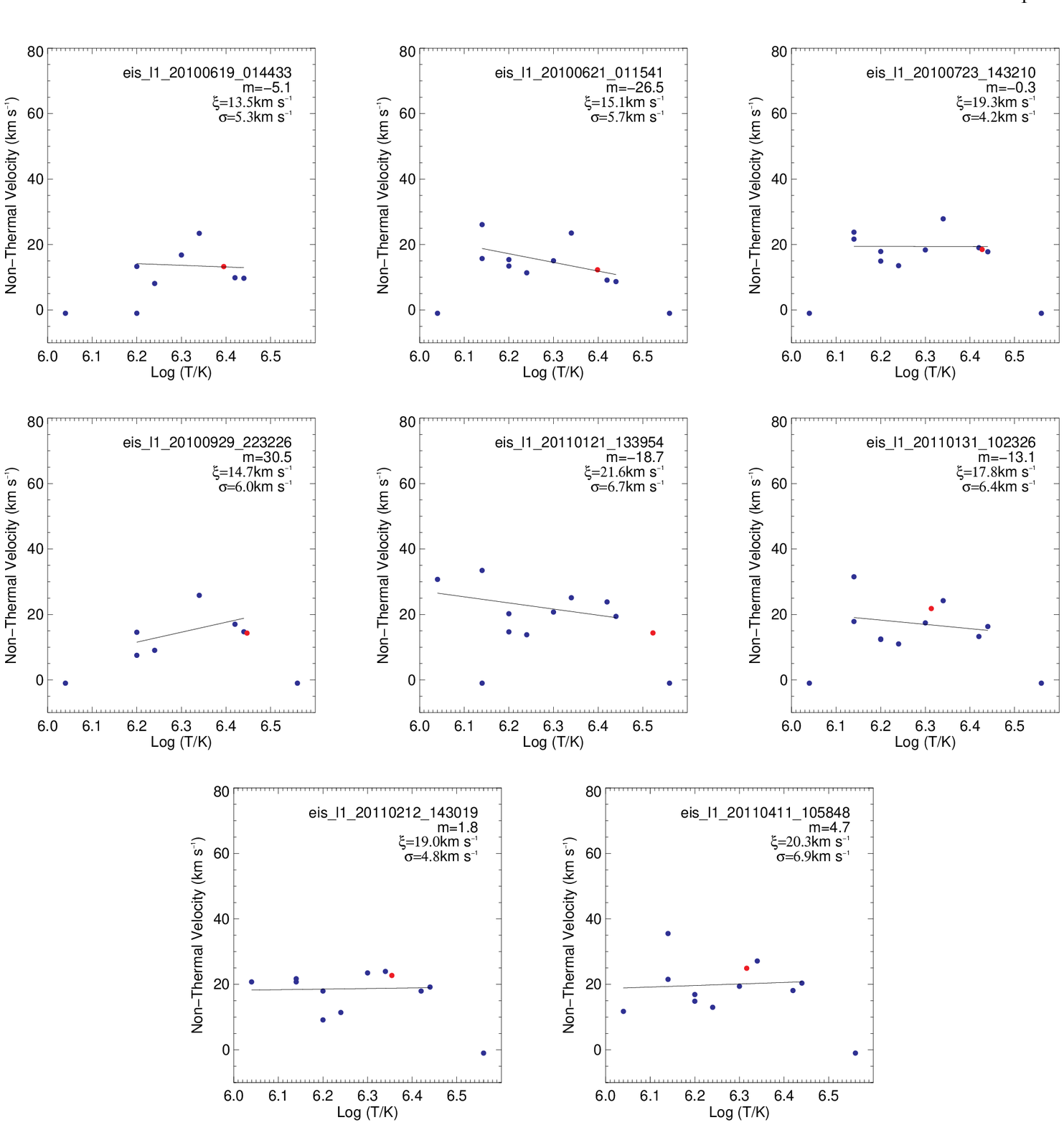}}
\caption{Plots of non-thermal velocity as a function of temperature for 8 of the active region
  cores. The datasets, gradient of the overlaid linear fit (m), mean non-thermal velocity ($\xi$),
  and standard deviation ($\sigma$) are indicated in the legend. The blue dots are the non-thermal
  velocities for each spectral line with an entry in Table \ref{table2} for the corresponding 
  dataset (cross-referenced in the Table and plot legend). The red dot is the non-thermal
  velocity calculated using the method of \citet{imada_etal2009} (see text). Data from all lines
  are included here.  }
\label{fig:fig5}
\end{figure*}

\begin{deluxetable}{rccccc}
\tabletypesize{\scriptsize}
\tablewidth{0pt}
\tablecaption{Active Region Observations}
\tablehead{
\multicolumn{1}{r}{\#} &
\multicolumn{1}{c}{EIS Dataset} &
\multicolumn{1}{c}{Slit\tablenotemark{a}} &
\multicolumn{1}{c}{FOV\tablenotemark{a}} &
\multicolumn{1}{c}{Exp.Time\tablenotemark{b}} &
\multicolumn{1}{c}{Study ID} \
}
\startdata
 1 & eis\_l1\_20100619\_014433 & 2 & 100x240 &  30 & 241\\
 2 & eis\_l1\_20100621\_011541 & 1 & 120x512 &  60 & 420\\
 3 & eis\_l1\_20100723\_143210 & 1 & 120x512 &  60 & 420\\
 4 & eis\_l1\_20100929\_223226 & 1 & 300x400 &  30 & 356\\
 5 & eis\_l1\_20110121\_133954 & 1 & 120x512 &  60 & 420\\
 6 & eis\_l1\_20110131\_102326 & 1 & 240x512 &  60 & 437\\
 7 & eis\_l1\_20110212\_143019 & 1 & 240x512 &  60 & 437\\
 8 & eis\_l1\_20110411\_105848 & 1 & 240x512 &  60 & 437\\
 9 & eis\_l1\_20110415\_001526 & 1 & 240x512 &  60 & 437\\
10 & eis\_l1\_20110419\_123027 & 1 & 240x512 &  60 & 437\\
11 & eis\_l1\_20110702\_030712 & 1 & 120x512 &  60 & 420\\
12 & eis\_l1\_20110725\_090513 & 1 & 120x512 &  60 & 420\\
13 & eis\_l1\_20110821\_105251 & 1 & 360x512 &  60 & 471\\
14 & eis\_l1\_20111108\_181234 & 1 & 240x512 &  60 & 437\\
15 & eis\_l1\_20111110\_100028 & 1 & 360x512 &  60 & 471  
\enddata
\tablenotetext{a}{Units of arcseconds.}
\tablenotetext{b}{Units are seconds.}
\label{table1}
\end{deluxetable}

\begin{figure*}[t!]
\centerline{\includegraphics[viewport= 40 215 585 790,clip,width=1.00\textwidth]{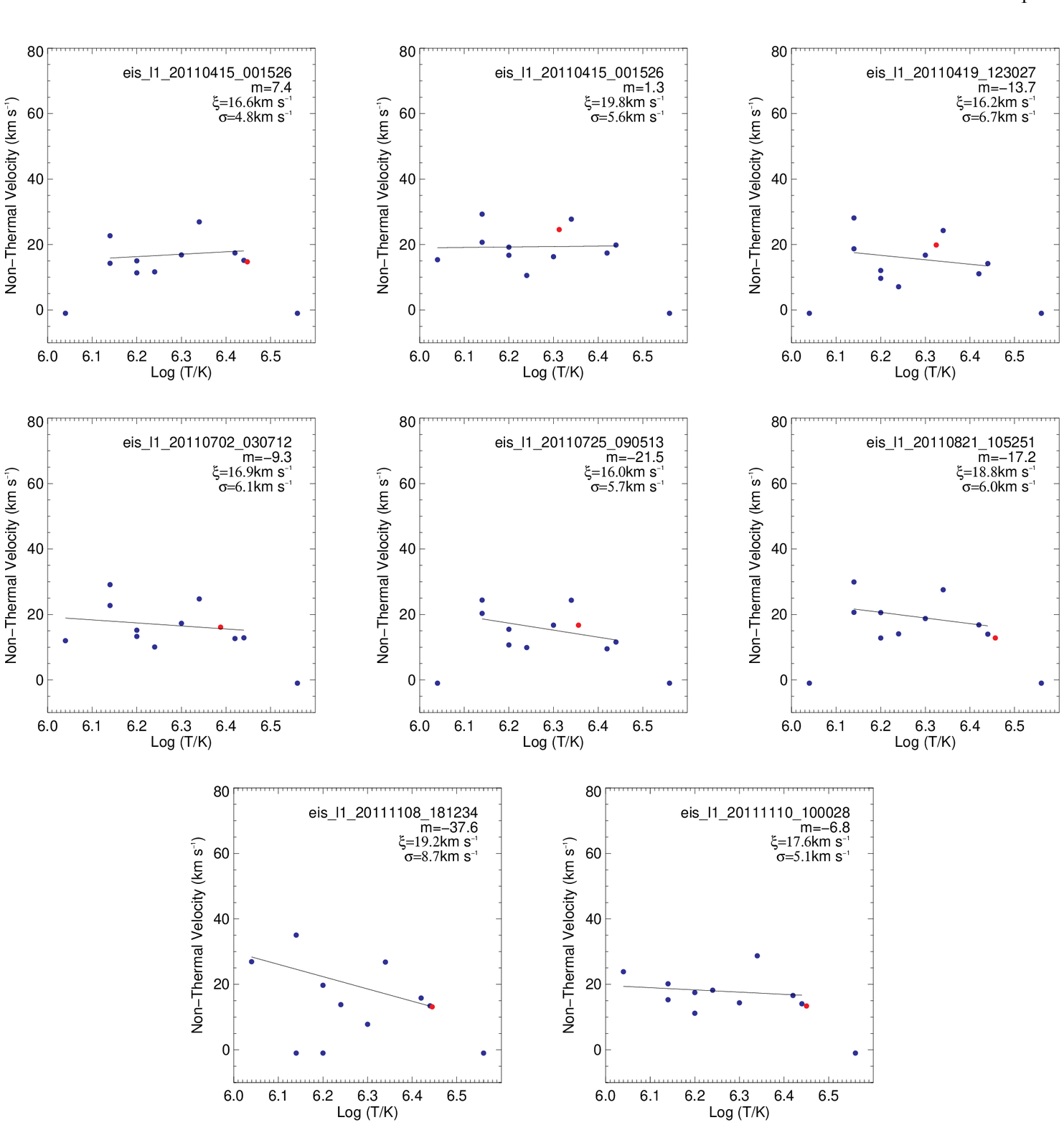}}
\caption{Plots of non-thermal velocity as a function of temperature for 8 of the active region
  cores. The datasets, gradient of the overlaid linear fit (m), mean non-thermal velocity ($\xi$),
  and standard deviation ($\sigma$) are indicated in the legend. The blue dots are the non-thermal
  velocities for each spectral line with an entry in Table \ref{table3} for the corresponding 
  dataset (cross-referenced in the Table and plot legend). The red dot is the non-thermal
  velocity calculated using the method of \citet{imada_etal2009} (see text). Data from all lines
  are included here.  }
\label{fig:fig6}
\end{figure*}

\begin{figure}[t!]
\centerline{\includegraphics[]{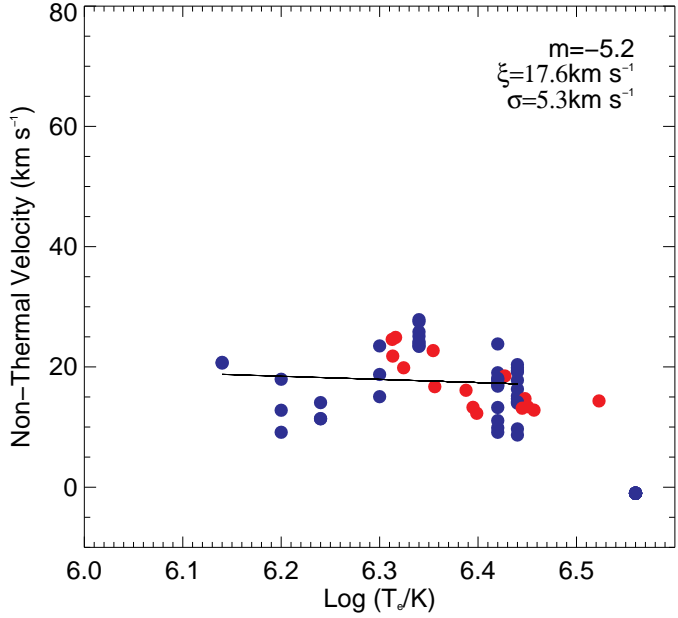}}
\caption{Non-thermal velocities as a function of line formation temperature for all 16 of the
  active region cores. Only lines with emission that is correlated with the emission of
  \ion{Ca}{14} 193.874\,\AA\, are included (see text). The quantities in the legend are the same
  as for Figures \ref{fig:fig5} and \ref{fig:fig6}, except that the non-thermal velocities
  calculated using the \citet{imada_etal2009} method are only shown for comparison and are not
  included in the average or standard deviation. These values are discussed in relation to Figure
  \ref{fig:fig8}.  }
\label{fig:fig7}
\end{figure}

\begin{figure}[t!]
\centerline{\includegraphics[]{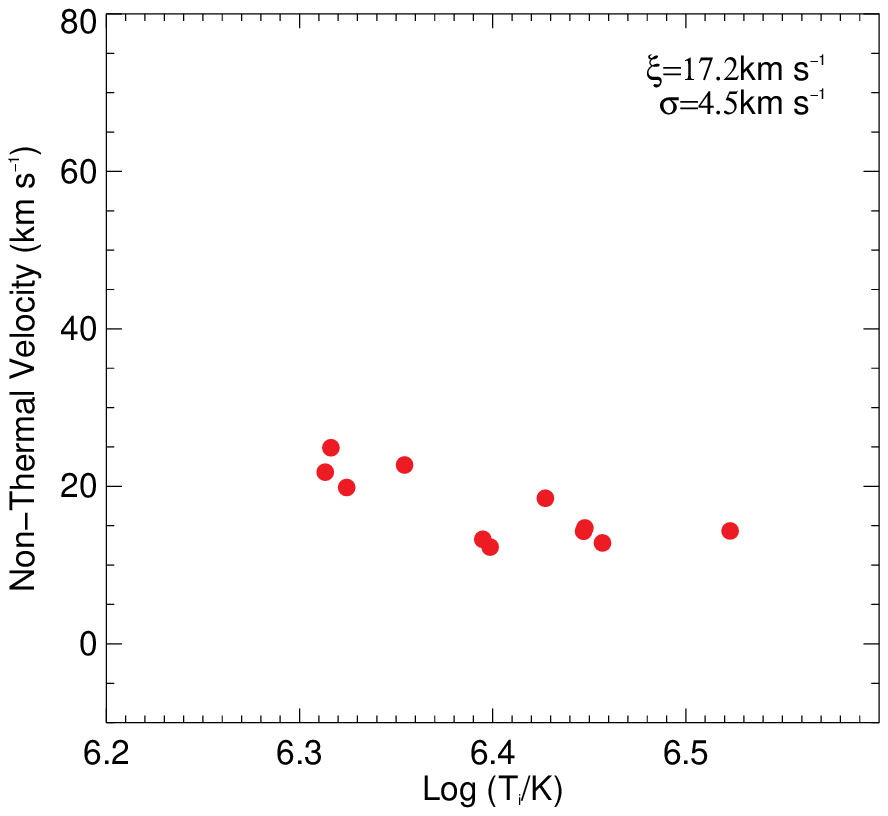}}
\caption{Non-thermal velocities calculated using the method of \citet{imada_etal2009} for all 16
  of the active region cores.  They are plotted as a function of the calculated ion
  temperature. Unlike Figures \ref{fig:fig5}, \ref{fig:fig6}, and \ref{fig:fig7}, the trend is not
  a function of temperature within an AR core. Rather it is a trend between different AR cores. To
  avoid confusion, we have not drawn a linear fit to these data.  }
\label{fig:fig8}
\end{figure}

For this study of spectral line widths, we use observations from EIS on {\it Hinode}
\citep{kosugi_etal2007}. EIS is a normal incidence spectrograph with multi-layer coatings on the
mirror and grating. It observes the solar EUV spectrum in two wavelength channels: a
short-wavelength (SW) channel covering 170--210\,\AA, and a long-wavelength (LW) channel covering
250--290\,\AA. This part of the spectrum is dominated by emission lines from the
\ion{Fe}{7}--\ion{Fe}{17} ionization stages of Iron, with numerous other lines from other
elements. These lines permit a very broad temperature coverage of observed solar features, and
many of the lines are also clean and relatively unblended, making them ideal for our purpose.
Lists of EIS spectral line identifications have been compiled by \citet{young_etal2007} and
\citet{brown_etal2008}.

EIS builds raster scan spectral images by stepping its slit across an area of the Sun using a scan
mirror. There are four slits: 1\arcsec, 2\arcsec, 40\arcsec, and 266\arcsec, with the narrowest
1\arcsec\ slit providing the best spectral resolution, which is about 22\,m\AA\, for 1 EIS
pixel. To obtain the most accurate line width measurements, we mostly use the highest spectral
resolution slit in this study (14/15 datasets). The choice of slit, field-of-view (FOV) of the
raster scan, spectral line list, exposure time etc. are fixed when the EIS observing study is
defined. We use data from five different EIS studies in this paper. For ease of reference, we list
the EIS datasets together with some pertinent information in Table \ref{table1}. The datasets are
the same as used by \citet{warren_etal2012} to study the emission measure (EM) distributions in
the hot AR core loops, so many details of the observations are also available in that paper.

EIS data are affected by a number of instrumental issues that need addressed prior to
analysis. There is a dark current pedestal that should be removed, and there are warm, hot, and
dusty pixels, and others that have been struck by cosmic-rays.  We removed these effects using the
eis\_prep routine in SolarSoftware (SSW, \citealt{freeland1998}). The centroid positions of the
spectral lines also show a periodic variation around the satellite orbit as a result of thermal
variations in the instrument structure.  We removed this effect using a neural network model
that relates velocity shifts to instrument temperatures \citep{kamio_etal2010}.

We show example spectral images for two of the datasets we analyzed in Figures \ref{fig:fig1} and
\ref{fig:fig2}.  We prepared these images by fitting single Gaussians to the observed spectra,
except for \ion{Fe}{12} 195.119\,\AA, where we used a double Gaussian fit to account for the
density sensitive blend in the red wing at 195.18\,\AA\, \citep{delzanna&mason_2003}. The figures
also indicate the boxed regions that we used for analysis of the hot core loops, and the fits
shown in the lower panels are example fits to the spectra averaged within these boxes.

Regarding our analysis strategy, it is important to note that we work with uncalibrated counts
(photons detected on the CCD) rather than absolute physical units derived from applying the
radiometric calibration.  Although the EIS effective areas are a slowly varying function of
wavelength \citep{lang_etal2006}, convolving them with the observed spectrum potentially
introduces another uncertainty into the analysis; especially since the EIS sensitivity has evolved
over time \citep{mariska_2013}, and there are different methods of characterizing that evolution
\citep{delzanna_2013,warren_etal2014}. Furthermore, the errors on the calibrated intensities also
depend on wavelength, and have an important bearing on the spectral fits. This can be significant,
depending on the treatment of the bad pixels. When we work in counts, the error is simply the
square root of the intensity, with no other complicating factors.

We demonstrate the effect of the absolute calibration quantitatively in Figure \ref{fig:fig3},
where we show scatter plot comparisons of the calibrated and uncalibrated line widths and
non-thermal velocities for the AR shown in Figure \ref{fig:fig1}. Although the calibrated and
uncalibrated line widths are highly correlated, some scatter is evident, and there is a tendency
for the calibrated widths to be larger, though we stress that this may depend on the dataset.
In our example, the calibrated widths are larger overall in about 90\% of the pixels. When we
eliminate the weakest pixels --- we use a threshold of 10\% of the maximum intensity in Figure
\ref{fig:fig1} --- the differences are $\sim$10-20\%. These differences, however, amplify when
propagated through to the non-thermal velocity computations. The scatter becomes larger, and
factor of 4.5 differences are possible.  In most cases, the differences are less than this extreme,
but a 10--20\% change in the average width for the \ion{Fe}{12} 192.394\,\AA\, line in this dataset
produces a 15--26\,km s$^{-1}$ change in non-thermal velocity. Furthermore, cases where the thermal
broadening is sufficient to explain the uncalibrated line width can produce erroneously large
values. These findings imply that previous measurements of line widths and non-thermal velocities
using calibrated EIS spectra should be treated with caution.  Avoiding these additional
uncertainties, together with the fact that for our study of line widths no absolute calibration is
necessary, is the reason we analyze the raw data prior to applying any absolute calibration.

Note that the non-thermal velocities shown in Figure \ref{fig:fig3} are calculated from
\begin{equation}
\delta\lambda = {\lambda_0 \over c} \sqrt{4\ln 2 ( {2k_BT_i \over m} + \xi^2 ) + \sigma_I^2 }
\label{eq:eq1}
\end{equation}
where $\delta\lambda$ is the observed line width, $\lambda_0$ is the line centroid, $k_B$ is
Boltzmann's constant, $T_i$ is the ion temperature, $m$ is the mass, $\xi$ is the non-thermal
velocity, and $\sigma_I$ is the instrumental width.

The instrumental width is a key parameter in the analysis. It was measured in the laboratory prior
to launch \citep{korendyke_etal2006} and the full width at half maximum (FWHM) was found to be
0.047\,\AA\, for the SW channel using the \ion{Mg}{3} 187\,\AA\, line and 0.055--0.057\,\AA\, for
the LW channel using \ion{He}{2} 256\,\AA\, and \ion{Ne}{3} 267\,\AA. \citet{brown_etal2008}
examined solar spectra acquired by \textit{Hinode} in orbit post-launch and found evidence that the SW instrumental width is broader 
(0.054\,\AA) than measured on the ground in the Rutherford Appleton Laboratory. \citet{hara_etal2011}
investigated the EIS instrumental width in detail by comparison of observed \ion{Fe}{14} lines
with ground based observations of the \ion{Fe}{14} 5303\,\AA\, green line obtained at the Norikura
Solar Observatory for an AR on the west limb in 2007, July 18.  They do not report values for the
SW channel, but their Figure 17 implies an average value of $\sim$0.062\,\AA\, for the LW
channel. They find only a gradual variation with wavelength, but more importantly, a significant
increase from $\sim$0.050\,\AA\, to $\sim$0.068\,\AA\, with Y-position on the CCD. This effect is
also discussed in detail by \citet{young_2011}, who modeled the Y-variation of the instrumental
width derived from the \ion{Fe}{12} 193.509\,\AA\, line in several off-limb quiet Sun datasets
obtained in 2009. \citet{young_2011} provided his correction as an IDL routine
(\verb+eis_slit_width+) in SSW, and we use this program to correct for the Y-variation in this
paper by constructing an array of pixel values using the Y-start position of the observations,
which is available in the EIS fits file header. In cases where we average over a small area, we
also average the Y-correction values.

The instrumental profile itself was found to be well represented by a Gaussian function in the
laboratory \citep{korendyke_etal2006} and we have independently verified this by re-examining the
data for \ion{Ne}{3} 267\,\AA. Deviations from a Gaussian profile may therefore be important
  signatures of the heating process. For example, numerical simulations of nanoflare heating
  predict some combination of Doppler shifted or highly distorted line profiles in the early phase,
  and subtle asymmetries later in the evolution \citep{patsourakos&klimchuk_2006}. Although EIS
line profiles have been found to often show important asymmetries \citep{hara_etal2008a}, that does
not seem to be the case for the fits to the profiles for the hot core loops, as can be seen in
Figures \ref{fig:fig1} and \ref{fig:fig2}. Nevertheless, since we are attempting to measure the
line widths as accurately as possible, we experimented with fits using a Voigt function with a
fixed damping factor based on a free-fit Gaussian function combined with a Lorentzian constructed
from the measured instrumental width. These fits were quite close to a Gaussian, but showed greater
dispersion, indicating that we would gain no advantage from moving to a more complex fitting
model. Furthermore, we closely examined the details of the Gaussian fits in Figures
  \ref{fig:fig1} and \ref{fig:fig2}, and we found that the total residual intensity within
  $\pm$100\,km s$^{-1}$ of the line centroids was always less than 5\% of the total intensity
  within the same wavelength range, indicating that the deviation from a Gaussian is small. This
  deviation also appears to decrease with temperature, making a Gaussian function even more
  appropriate for the highest temperature lines which are of the most interest for this study. 
For all of these reasons, and also for ease of comparison with the literature results, we use only
a simple Gaussian function for the fits in this paper. Note that we also use relatively narrow
spectral windows for our fits. This is to avoid interference from background in the wings of the
profiles (see the discussion related to \ion{Ca}{14} 193.874\,\AA\, below).

Ideally we would use only clean, unblended lines to obtain the best line width measurements, but
since we are interested in any trend as a function of temperature, we have to compromise in some
cases. Furthermore, most EIS datasets contain a limited number of spectral lines in order to
conserve telemetry, so in some cases we have had to choose additional lines that we otherwise
would not use. The lines we selected are given in Tables \ref{table2} and \ref{table3}, and some
more detailed comments on the reasons for selection are appropriate here.

First, \ion{Fe}{10} 184.526\,\AA, \ion{Fe}{11} 180.401\,\AA, \ion{Fe}{12} 192.394\,\AA,
\ion{Fe}{13} 202.044\,\AA, \ion{Fe}{14} 264.787\,\AA, \ion{Fe}{16} 262.984\,\AA, and \ion{Ca}{14}
193.874\,\AA\, are all clean, unblended and are either the strongest line for their ionization
stage or have sufficient signal to produce good spectral fits. They cover the temperature range
from 1.1--3.6\,MK. \ion{Si}{10} 258.275\,\AA\, is also unblended and we include it in order to
compare the \ion{Si}{10} and \ion{Fe}{12} widths. Second, since not all of our datasets include
the \ion{Fe}{14} 264.787\,\AA\, line, we added \ion{Fe}{14} 274.203\,\AA\, as a backup choice. We
also include \ion{Fe}{15} 284.160\,\AA. This is a strong line and fills the temperature gap
between \ion{Fe}{14} and \ion{Fe}{16}, which is otherwise quite large. We also include \ion{S}{13}
256.686\,\AA\, for comparison with \ion{Fe}{16} 262.984\,\AA. These last three lines are weakly
blended, \ion{Fe}{14} 274.203\,\AA\, with \ion{Si}{7} 274.175\,\AA\, and \ion{Fe}{15}
284.160\,\AA\, with \ion{Al}{9} 284.015\,\AA\, and \ion{S}{13} 256.686\,\AA\, with \ion{Ni}{16}
256.62\,\AA, but we expect any systematic deviation in line width to become apparent in comparison
with the other lines.

\begin{deluxetable*}{lcccccccc}
\tabletypesize{\scriptsize}
\tablewidth{0pt}
\tablecaption{Line FWHMs and Non-Thermal Velocities}
\tablehead{
\multicolumn{8}{c}{Active Region Observation Date} & \\
[.3ex]\cline{2-9} \\[-1.6ex] 
\multicolumn{1}{l}{Line ID} &
\multicolumn{1}{r}{ 20100619} &
\multicolumn{1}{r}{ 20100621} &
\multicolumn{1}{r}{ 20100723} &
\multicolumn{1}{r}{ 20100929} &
\multicolumn{1}{r}{ 20110121} &
\multicolumn{1}{r}{ 20110131} &
\multicolumn{1}{r}{ 20110212} &
\multicolumn{1}{r}{ 20110411} \
}
\startdata
   Fe X 184.536   & 0.069 [0.00]& 0.061 [0.00]& 0.062 [0.00]& 0.062 [0.00]& 0.069 [30.7]& 0.062 [0.00]& 0.065 [20.7]& 0.063 [11.7] \\
  Fe XI 180.401         & --    & 0.068 [26.0]& 0.067 [23.7]      & --    & 0.071 [33.4]& 0.070 [31.5]& 0.066 [21.7]& 0.072 [35.5] \\
 Fe XII 192.394   & 0.073 [13.3]& 0.066 [13.4]& 0.067 [17.8]& 0.066 [14.5]& 0.067 [20.2]& 0.065 [12.4]& 0.066 [17.9]& 0.066 [16.8] \\
 Fe XII 195.119   & 0.071 [0.00]& 0.066 [15.3]& 0.066 [14.9]& 0.065 [7.51]& 0.065 [14.6]& 0.065 [12.3]& 0.064 [9.13]& 0.066 [14.8] \\
Fe XIII 202.044   & 0.073 [8.10]& 0.066 [11.3]& 0.067 [13.5]& 0.066 [9.03]& 0.066 [13.8]& 0.066 [10.9]& 0.065 [11.4]& 0.066 [12.9] \\
 Fe XIV 264.787         & --    & 0.073 [15.0]& 0.075 [18.3]      & --    & 0.075 [20.7]& 0.074 [17.4]& 0.077 [23.5]& 0.075 [19.4] \\
 Fe XIV 274.203   & 0.081 [16.8]      & --          & --          & --          & --          & --          & --          & --     \\
  Fe XV 284.160   & 0.087 [23.4]& 0.081 [23.5]& 0.084 [27.8]& 0.083 [25.8]& 0.081 [25.1]& 0.081 [24.2]& 0.080 [23.9]& 0.084 [27.1] \\
 Fe XVI 262.984   & 0.081 [9.70]& 0.074 [8.68]& 0.077 [17.7]& 0.076 [14.7]& 0.077 [19.4]& 0.077 [16.3]& 0.077 [19.1]& 0.078 [20.3] \\
   Si X 258.375         & --    & 0.076 [15.7]& 0.079 [21.6]      & --    & 0.067 [0.00]& 0.077 [17.8]& 0.077 [20.7]& 0.078 [21.5] \\
 S XIII 256.686   & 0.087 [9.86]& 0.081 [9.13]& 0.084 [19.0]& 0.083 [17.0]& 0.086 [23.8]& 0.082 [13.2]& 0.083 [17.9]& 0.083 [18.1] \\
 Ca XIV 193.874         & --    & 0.066 [0.00]& 0.072 [0.00]& 0.069 [0.00]& 0.069 [0.00]& 0.072 [0.00]& 0.065 [0.00]& 0.070 [0.00] \\
          $\xi$           & 13.2        & 12.2        & 18.4        & 14.3        & 14.3        & 21.8        & 22.7        & 24.9  \
\enddata
\tablenotetext{*}{The FWHMs are in units of \,\AA\, and the non-thermal velocities
are in km s$^{-1}$. $\xi$ is the non-thermal velocity calculated using the \citet{imada_etal2009} method.
\label{table2}}
\end{deluxetable*}

\begin{deluxetable*}{lcccccccc}
\tabletypesize{\scriptsize}
\tablewidth{0pt}
\tablecaption{Line FWHMs and Non-Thermal Velocities}
\tablehead{
\multicolumn{8}{c}{Active Region Observation Date} & \\
[.3ex]\cline{2-9} \\[-1.6ex] 
\multicolumn{1}{l}{Line ID} &
\multicolumn{1}{r}{ 20110415} &
\multicolumn{1}{r}{ 20110415} &
\multicolumn{1}{r}{ 20110419} &
\multicolumn{1}{r}{ 20110702} &
\multicolumn{1}{r}{ 20110725} &
\multicolumn{1}{r}{ 20110821} &
\multicolumn{1}{r}{ 20111108} &
\multicolumn{1}{r}{ 20111110} \
}
\startdata
   Fe X 184.536   & 0.062 [0.00]& 0.064 [15.3]& 0.059 [0.00]& 0.063 [11.9]& 0.063 [0.00]& 0.061 [0.00]& 0.071 [26.9]& 0.069 [23.8] \\
  Fe XI 180.401   & 0.067 [22.6]& 0.069 [29.2]& 0.069 [28.1]& 0.069 [29.1]& 0.068 [24.3]& 0.069 [29.9]& 0.075 [35.0]& 0.068 [20.1] \\
 Fe XII 192.394   & 0.066 [15.0]& 0.067 [19.2]& 0.065 [12.0]& 0.065 [15.1]& 0.067 [15.4]& 0.067 [20.5]& 0.070 [19.7]& 0.069 [17.4] \\
 Fe XII 195.119   & 0.066 [11.3]& 0.066 [16.7]& 0.065 [9.67]& 0.065 [13.2]& 0.066 [10.6]& 0.065 [12.8]& 0.066 [0.00]& 0.067 [11.1] \\
Fe XIII 202.044   & 0.067 [11.6]& 0.066 [10.5]& 0.066 [7.11]& 0.065 [10.0]& 0.067 [9.88]& 0.066 [14.0]& 0.069 [13.8]& 0.070 [18.2] \\
 Fe XIV 264.787   & 0.074 [16.8]& 0.073 [16.2]& 0.074 [16.7]& 0.074 [17.2]& 0.075 [16.7]& 0.074 [18.7]& 0.073 [7.80]& 0.075 [14.3] \\
  Fe XV 284.160   & 0.084 [26.9]& 0.084 [27.7]& 0.082 [24.2]& 0.081 [24.7]& 0.082 [24.3]& 0.084 [27.5]& 0.086 [26.8]& 0.087 [28.7] \\
 Fe XVI 262.984   & 0.076 [15.1]& 0.078 [19.8]& 0.076 [14.1]& 0.075 [12.8]& 0.076 [11.5]& 0.075 [13.9]& 0.078 [13.4]& 0.078 [14.0] \\
   Si X 258.375   & 0.076 [14.2]& 0.078 [20.6]& 0.077 [18.7]& 0.079 [22.7]& 0.079 [20.3]& 0.078 [20.6]& 0.070 [0.00]& 0.078 [15.3] \\
 S XIII 256.686   & 0.084 [17.4]& 0.083 [17.4]& 0.081 [11.0]& 0.081 [12.6]& 0.081 [9.50]& 0.083 [16.8]& 0.085 [15.8]& 0.085 [16.6] \\
 Ca XIV 193.874   & 0.070 [0.00]& 0.070 [0.00]& 0.069 [0.00]& 0.067 [0.00]& 0.064 [0.00]& 0.069 [0.00]& 0.071 [0.00]& 0.072 [0.00] \\
          $\xi$           & 14.7        & 24.5        & 19.8        & 16.1        & 16.7        & 12.8        & 13.1        & 13.3  \
\enddata
\tablenotetext{*}{The FWHMs are in units of \,\AA\, and the non-thermal velocities
are in km s$^{-1}$. $\xi$ is the non-thermal velocity calculated using the \citet{imada_etal2009} method.
\label{table3}}
\end{deluxetable*}

In addition to computing non-thermal velocities using Equation \ref{eq:eq1}, and as an independent
check, we also calculate them using a different method outlined by \citet{imada_etal2009}
(also see \citealt{seely1997}). \citet{imada_etal2009} used lines from ions of
different masses, and, under the assumption that the non-thermal velocity and ion temperature are
the same, derived expressions from which they could be computed.

\begin{equation}
\xi^2 = {m_2 W_2^2 - m_1 W_1^2 \over 4\log 2 (m_2 - m_1)}
\label{eq:eq2}
\end{equation}
and
\begin{equation}
T_{ion} = {W_1^2 - W_2^2 \over 8k_B \log 2} {m_1 m_2 \over m_2 - m_1}
\label{eq:eq3}
\end{equation}
where $W=\delta\lambda^2-\sigma_{I}^2$, and $1$ and $2$ indicate the ion species.
\citet{imada_etal2009} used \ion{S}{13} 256.686\,\AA\, and \ion{Fe}{16} 262.984\,\AA\, for their
study, and we use the same line pair here.

We analyzed 16 high temperature core loop arcades from the 15 datasets listed in table
\ref{table1}. Figure \ref{fig:fig4} shows \ion{Fe}{18} 93.92\,\AA\, images of the ARs that were
extracted from AIA 94\,\AA\, filtergrams using the technique described by
\citet{warren_etal2012}. The green boxes indicate the regions where the measurements were made.
These regions were carefully selected to isolate the loops that are bright in \ion{Fe}{18}
93.92\,\AA, with minimal contamination from footpoint (moss) emission below. The AIA 193\,\AA\,
filtergrams were then coaligned with the EIS \ion{Fe}{12} 195.119\,\AA\, raster images, and the
relevant areas extracted for all of the lines of interest. More details of the selection and
coalignment methods are given in \citet{warren_etal2012}.

The results of our measurements are displayed in Figures \ref{fig:fig5}--\ref{fig:fig6} and the
values are noted explicitly in Tables \ref{table2}--\ref{table3}. Figures \ref{fig:fig5} and
\ref{fig:fig6} show the measured non-thermal velocities plotted as a function of temperature for
the 16 core loop arcades. The non-thermal velocities are all less than 35\,km s$^{-1}$ with mean
values in the range 13.5--21.6\,km s$^{-1}$ with a standard deviation of 4.2--8.7\,km s$^{-1}$.
Linear fits to the non-zero measurements are overlaid to draw the eye to any trend in the data,
but taking into consideration the scatter in the measurements, there is no statistically
significant trend.

We present the complete results in the figures and tables so that they are on record, but it is
not necessarily the case that the emission at all temperatures comes from the hot core loops. If
that were the case we would expect these loops to have broad EMs whereas \citet{warren_etal2012}
found strongly peaked EM distributions with a sharp fall below and above the peak temperature in
most cases. In some of the weaker ARs a shallower temperature distribution was observed, and this
highlights the need to adopt some method of quantitatively determining if the emission comes from
the hot core loops or not. In our previous EM studies of warm coronal loops
\citep{warren_etal2008b,brooks_etal2012,brooks_etal2013} we have adopted the criterion that the
cross-field intensity profile must be highly correlated with that of the spectral line with which
they were identified (usually \ion{Fe}{12} 195.119\,\AA). Here we impose the condition that the
emission within the boxed region for each spectral line must be spatially correlated ($r>0.6$) with that of the highest temperature line (\ion{Ca}{14} 193.874\,\AA).  The measurements satisfying this criterion for all the AR loops studied are shown in Figure \ref{fig:fig7}.  Note that no lines formed at temperatures below 1.4\,MK pass this test.

Returning to the measurements, the mean value for all the hot loops is 17.6\,km s$^{-1}$ with a
standard deviation of 5.3\,km s$^{-1}$, and as with the individual plots (Figures
\ref{fig:fig5}--\ref{fig:fig6}) the two methods of non-thermal velocity calculation appear to give
fairly consistent results.

Note that the highest temperature \ion{Ca}{14} 193.874\,\AA\, measurements are always zero. This
is inconsistent with the results for the other highest temperature line, \ion{Fe}{16}
262.984\,\AA, and also with previous ground based measurements of \ion{Ca}{15} 5694\,\AA\,
obtained with the Norikura solar observatory coronagraph by \citet{hara&ichimoto_1999}. The
Norikura coronagraph has superior spectral resolving power to EIS and obtained values of $\sim$
16--26\,km s$^{-1}$ for a limb AR, which would be fairly consistent with our other
results. Something is therefore missing in our understanding of the EIS \ion{Ca}{14}
193.874\,\AA\, line width measurements.

One possibility is that the \ion{Ca}{14} 193.874\,\AA\, measurement can be attributed to an
unusual deviation of the instrumental width. It lies very close to the \ion{Fe}{12} 192.394\,\AA\,
line which shows similar values for the measured line width --- within 7\%; see Tables \ref{table2}
and \ref{table3} --- and the \ion{Ca}{14} 193.874\,\AA\, non-thermal velocities could be brought
into agreement with those of \ion{Fe}{12} 192.394\,\AA\, if the instrumental width drops to
0.047\,\AA\, at 193.874\,\AA. This would actually be in agreement with the SW laboratory
measurement of the instrumental width, but in disagreement with most on orbit measurements.  It is
difficult to understand how the laboratory calibration could be maintained at one wavelength while
degrading by $\sim$ 40\% within 1.5\,\AA, however, so we consider this explanation unlikely.

Another possibility is that \ion{Ca}{14} 193.874\,\AA\, is formed at a much lower temperature than
suggested by the ionization equilibrium calculations. The contribution function peaks at 3.6\,MK,
but if the line were actually formed below 1\,MK the non-thermal velocity measurements would be
non-zero. We also consider this explanation unlikely since the EM distributions for all of these
regions actually peak close to the equilibrium formation temperature of \ion{Ca}{14}
193.874\,\AA\, \citep{warren_etal2012}.

Another explanation for this observation is that the complex background in this region of the
spectrum adversely affects the line profile fits. Since the background is high, sloped, and
affected by several weaker lines, there is a possibility that the full height of the line profile
is not exposed, and therefore that the width measurement is made slightly higher than the half
maximum location. This issue can actually resolve the discrepancy between the \ion{Ca}{14}
193.874\,\AA\, non-thermal velocity measurements and the others, bringing the non-thermal
velocities up to values of 19--33\,km s$^{-1}$. We have also independently verified that we can
measure non-zero non-thermal velocities in the \ion{Ca}{14} 193.874\,\AA\, line during an X1.8
class solar flare when the line is more prominently exposed above the background.  Since a
definitive proof that this is the correct explanation requires higher spectral resolution
observations of this line, however, we discuss it only in the Appendix.

Figure \ref{fig:fig8} shows the non-thermal velocity results calculated using Equation
\ref{eq:eq2} and plotted against the ion formation temperature calculated using Equation
\ref{eq:eq3} for the AR core loops where the emission from the diagnostic \ion{S}{13}
256.686\,\AA\, and \ion{Fe}{16} 262.984\,\AA\, lines is correlated with \ion{Ca}{14}
193.874\,\AA. The average value in this case is 17.2\,km s$^{-1}$ with a standard deviation of
4.5\,km s$^{-1}$, which is consistent with the previous results shown.  The ion temperatures fall in
the range 2--3\,MK, which are slightly lower than the temperature of the peak of the EM calculated
by \citet{warren_etal2012}, but not low enough to solve the \ion{Ca}{14} 193.874\,\AA\,
non-thermal velocity discrepancy. The plot also shows a decreasing non-thermal velocity with
increasing ion formation temperature of each AR. Note that this is a trend between ARs not within
any single one.

\citet{warren_etal2012} provide other parameters for this sample of AR core loops in their Table 1
such as the total unsigned magnetic flux of the AR, the total AIA \ion{Fe}{18} intensity, and the
gradient of the EM slope below and above the peak. We have verified that there is no correlation
between any of these parameters and our measured non-thermal velocities.

Finally, we also investigated any relationship between the non-thermal velocities measured
in \ion{Fe}{16} 262.984\,\AA\, and the loop lengths, by attempting to trace the loops from
the foot-point to the apex in the \ion{Fe}{18} 93.92\,\AA\, images of Figure \ref{fig:fig4}.
The results are not always directly comparable: in most cases the tracing was satisfactory, but sometimes
only part of the loop length was visible, or the boxed region where the non-thermal velocities were measured
caught parts of other hot loops in the AR core; in which case we measured the length of another loop
in the AR core arcade. While there were potentially interesting trends in some subsets of the data, in 
general we found the correlation between loop length and non-thermal velocity to be poor. 

\section{summary and discussion}

We have carried out a survey of non-thermal line widths in the high temperature (1.1--3.6\,MK)
loops in the cores of 15 non-flaring ARs spanning a wide range of solar conditions. We compute the
non-thermal velocities considering the instrumental and thermal broadening of the spectral lines
in the usual way, and we also utilize the method outlined by \citet{imada_etal2009}.  The results
from the two methods are broadly in agreement, and we find non-thermal velocities of $\sim$ 17\,km
s$^{-1}$ on average.  We also find no significant trend with temperature, or with any other
property of the ARs such as total magnetic flux, or the slope of the loop EM distributions. We do,
however, detect a tendency for AR loops with the highest ion temperatures to have smaller
non-thermal velocities.

We note that stellar observers have concluded that non-thermal velocities below $\sim$\,20\,km
s$^{-1}$ indicate that the spectral lines are basically thermally broadened
\citep{linsky_etal1998}. While that is not strictly the case here, the modest values we find, and 
the lack of a temperature trend, are a
challenge for current coronal heating models to explain. 
These are two of the key measurements that in combination can be used not just for discriminating between different coronal
heating models, but also for distinguishing between the physical processes within the same model.
For example, in the nanoflare model, reconnection jets are expected to exist while the plasma is still
too hot to have cooled sufficiently for emission to appear at lower temperatures. So a comparison with
the non-thermal velocity dependence on temperature is not appropriate. Conversely, emission
at lower temperatures is dominated by the later phase after the nanoflare has ended and
such a comparison becomes possible. We find that our measured non-thermal velocities 
are much smaller than predicted from
either the high temperature reconnection jets in the nanoflare-heated corona model, or shock
heating associated with Alfv\'{e}n waves, both of which suggest velocities on the order of
hundreds of km s$^{-1}$ \citep{cargill_1996,antolin_etal2008}. Models of chromospheric evaporation
in response to coronal nanoflares are closer to the observations around 1\,MK, but 
predict that non-thermal velocities should increase with temperature. This is because the emission
at cooler temperatures is low while chromospheric evaporation is occurring and is dominated by the later
slow draining phase, whereas the emission at higher temperatures occurs earlier in the plasma evolution and 
is produced not just by draining but also by evaporation. This trend, however, is not observed, and 
leads to a discrepancy of a factor of 2 around 4\,MK \citep{patsourakos&klimchuk_2006}.  Some
models of Alfv\'{e}n wave turbulence also produce values that are closer to the observations
around 1.6\,MK, and show only limited variations with temperature at the loop apex, but they show 
a large spread of predictions, and require the imposition of a
random flow component parallel to the magnetic field \citep{asgaritarghi_etal2014}.

Of course, further modeling with different physical conditions may alter the predicted characteristics 
of the spectra. For example, non-equilibrium ionization will alter the line intensities,
and this effect was recently demonstrated to be
significant in the case of flare-like reconnection jets \citep{imada_etal2011}. A change in the line intensity
implies a potential alteration of the line width. It seems more intuitive to think that a more
violent plasma will produce more dynamic spectral signatures, but it is possible that this is not
the case, and it should be checked with numerical simulations. Temperature trends could also be
affected. 

For the nanoflare model it is possible that high temperature reconnection jets exist, but we are
unable to observe them in AR core loops because the EM is too low (as discussed earlier). Non-equilibrium ionization
could contribute to surpressing the emission too \citep{bradshaw&cargill_2006}. In this picture,
the non-thermal broadening of the spectral lines would appear as it does in real flares,
with the hot AR core loops corresponding to post-flare loops observed in their cooling phase. In
fact, EIS observations of flares do sometimes show substantial line broadening at high
temperatures. \citet{hara_etal2008b} observed a long-duration event on the limb, and measured
significant non-thermal velocities of 125\,km s$^{-1}$ at the top of cusp-shaped flare loops in
the \ion{Ca}{17} 192.858\,\AA\, line formed near 5.6\,MK.  \citet{hara_etal2008b} and
\citet{doschek_etal2014}, both noted that the non-thermal velocities at coronal temperatures in
flares (1.1--2.2\,MK) tend to be smaller (10--40\,km s$^{-1}$), which is consistent with the
velocities diminishing as flare loops cool. \citet{hara_etal2008b} actually made the measurements
in post-flare loops on the limb. 
 
These results would appear to fit the idea that the heating of AR core loops is truly flare-like,
but we only observe the non-thermal broadening that remains well after the energy release process
is complete. In this sense, measurements of non-thermal broadening are not a good diagnostic of
the heating phase. In fact there have also been studies showing tentative signals of cooling
plasma in the hot cores of ARs \citep{viall&klimchuk_2012}.

The picture for flares, however, is incomplete. 
\citet{doschek_etal2014}, for example, only measured values
of 20--60\,km s$^{-1}$ in \ion{Fe}{24} in their sample of M- and X-class flares.
Furthermore, there is a body of evidence that suggests that the hot loops in the cores of ARs 
are not cooling, but instead are maintained at high temperatures by continual high frequency 
heating \citep{antiochos_etal2003,brooks&warren_2009,warren_etal2010}. 

This work cannot draw a firm conclusion on the heating process, but we present a systematic
observational study of non-thermal broadening in high temperature core loops that provides
an important observational constraint for coronal heating models.


\acknowledgments The authors would like to thank Peter Cargill for suggesting this project. This
work was funded by the NASA {\it Hinode} program. {\it Hinode} is a Japanese mission developed and
launched by ISAS/JAXA, with NAOJ as domestic partner and NASA and STFC (UK) as international
partners. It is operated by these agencies in co-operation with ESA and NSC (Norway).

\appendix
\section{background influence on the \ion{Ca}{14} 193.874\,\AA\, line profile}

\begin{figure*}[t!]
\centerline{\includegraphics[width=\textwidth]{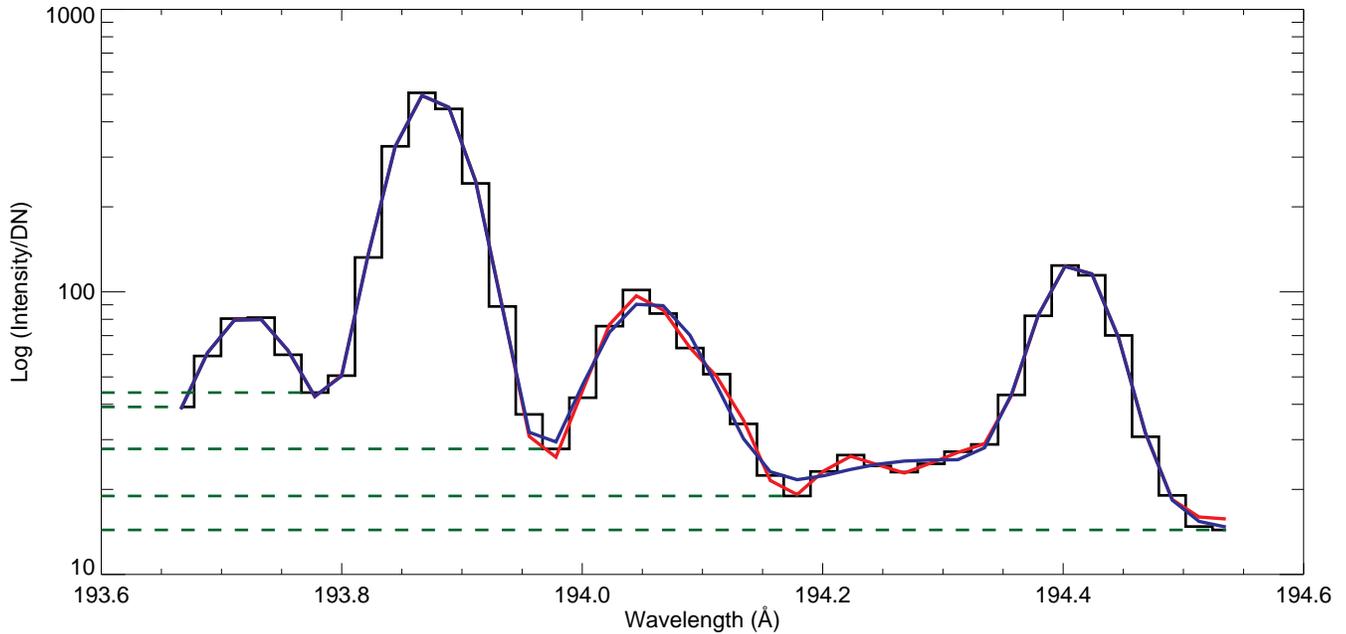}}
\caption{Sample spectrum close to \ion{Ca}{14} 193.874\,\AA\, (black histogram). The data are of the
hot AR core loop observed on 2010, September 29. The blue and red lines show example fits using
different locations for a linear fit to the background level. The horizontal dashed green lines
indicate the variation in background level depending on the spectral pixel where it is
measured. The intensity scale is logarithmic to highlight the variation in background.  There are
seven spectral lines within this wavelength range (see text).}
\label{fig:fig9}
\end{figure*}

The spectral interval immediately surrounding the \ion{Ca}{14} 193.874\,\AA\, line is fairly
crowded, with the following lines present according to \citet{brown_etal2008}:
\ion{Fe}{10} 193.715\,\AA, \ion{Fe}{8} 193.967\,\AA, a blend of \ion{Ni}{16} 194.046\,\AA\, and 
\ion{Ar}{11} 194.104\,\AA, and 
\ion{Ar}{14} 194.396\,\AA. There are also unidentified lines at 
194.227\,\AA\, and 194.315\,\AA. The background level is also sloped and difficult to determine,
and as discussed in the main paper, these effects may partially conceal the \ion{Ca}{14}
193.874\,\AA\, line profile and result in the width measurement at half-maximum intensity actually
being taken slightly above the half maximum. This is illustrated in Figure \ref{fig:fig9}, which
shows a sample spectrum in this wavelength region. The blue and red fits to the data (black
histogram) are made by selecting slightly different points as initial guesses for a linear
background fit in order to test cases where the background is either well matched close to or far
from the \ion{Ca}{14} 193.874\,\AA\, line profile. Note that the intensity scale is logarithmic in
order to highlight the varying background levels, so the differences in these fits are small, and
the width measurements for \ion{Ca}{14} 193.874\,\AA\, are not substantially affected.

The background level near \ion{Ar}{14} 194.396\,\AA, however, is $\sim$2 times smaller than the
background level on the long wavelength side of \ion{Ca}{14} 193.874\,\AA\, and $\sim$3 times
smaller than on the short wavelength side. The variation of the background depending on where it
is measured is illustrated by the horizontal green lines. The template fits for the results in the
main part of the paper measure the background close to the line profile where it is relatively
high, however, the actual measured width as a result of the more elaborate fit shown in
Figure \ref{fig:fig9} is not significantly different, 0.031\,\AA\, Gaussian width compared to
0.030\,\AA, and cannot resolve the discrepancy in non-thermal velocities.  Conversely, the base of
the \ion{Ca}{14} 193.874\,\AA\, line appears partly hidden by the sloping background and blending
with the wings of the \ion{Fe}{10} 193.715\,\AA\, and \ion{Fe}{8} 193.967\,\AA\, lines.

\begin{figure*}[t!]
\centerline{\includegraphics[width=\textwidth]{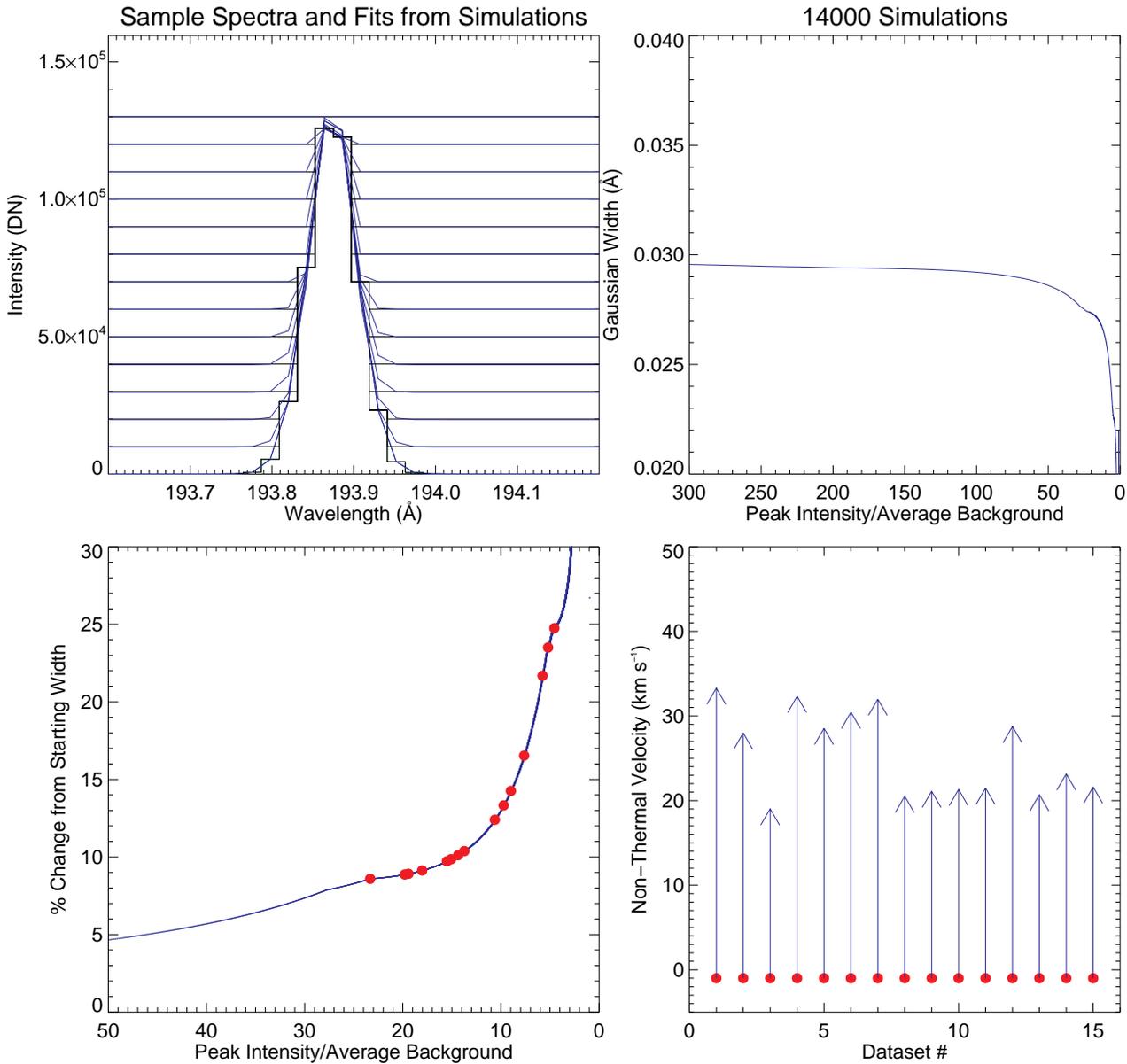}}
 \caption{Simulations of the effect of enhanced background on the \ion{Ca}{14} 193.874\,\AA\,
spectral line profile. Top left: a random sample of simulated profiles with increasing background
levels. Gaussian fits to the profiles are overlaid in blue. Top right: The Gaussian width of the
line as a function of the ratio of the peak intensity to average background level. Bottom left:
the percentage change in the line profile width (compared to the initial width) as a function of
the ratio of the peak intensity to average background level. The data for the 16 AR core loops are
overplotted as red dots at the actual measured values of the ratio. Bottom right: arrows showing
the adjusted non-thermal velocities if the simulated correction were applied to the real data.  }
\label{fig:fig10}
\end{figure*}

Without higher spectral resolution observations we cannot confirm that the profile is affected,
but we have performed a series of experiments to simulate what the effect would be. These are
illustrated in Figure \ref{fig:fig10}. The top left panel shows a simulated \ion{Ca}{14}
193.874\,\AA\, line profile (black histogram) with Gaussian fits to randomly selected profiles
with successively higher background levels in blue. The initial profile is constructed with a
Gaussian area of 10000\,DN so that the intensity error is less than 1\% of the total intensity. The
Gaussian line width was chosen as 0.030\,\AA.  This profile is then fit with a Gaussian function
plus linear background and the background fit extracted.  The background is then increased,
re-convolved with the original Gaussian part, and a further Gaussian fit is made. This process is
repeated until the Gaussian part of the function is completely swamped by the background.

We extracted the line width from the Gaussian fits to each simulated profile, and plotted them as
a function of the profile peak intensity to average background ratio in the top right panel of
Figure \ref{fig:fig10}. We can see that the line width is a slowly decreasing function of the
ratio, but decreases rapidly as the Gaussian part of the profile disappears. To examine the
significance of the effect, we plotted the percentage change in the line width (compared to the
starting value) as a function of the same profile peak intensity to average background ratio in
the lower left panel of Figure \ref{fig:fig10}. The width changes by less than 10\% until the
ratio is larger than $\sim$15, but we know from Figure \ref{fig:fig3} in the main paper that line
width changes of that order can result in significant changes in computed non-thermal
velocities. So we have overlaid red dots on the figure at the locations of the actual measured
profile peak intensity to average background ratio for the 16 AR core loops in our study, to
determine how much each of our actual measurements would be affected. From the Figure it appears
that our line width measurements could change by 9-25\%, which is significant.

The lower right panel of Figure \ref{fig:fig10} shows arrows indicating the change in computed
non-thermal velocity for the sample of \ion{Ca}{14} 193.874\,\AA\, measurements as a result of the
percentage changes indicated by the lower left panel. Our experiments imply that when the
background effect is taken into account, the revised \ion{Ca}{14} 193.874\,\AA\, non-thermal
velocities for the sample AR core loops fall in the range 19--33\,km s$^{-1}$ with an average value
of 25\,km s$^{-1}$. This is broadly in agreement with the results for the other lines within the
uncertainties. As noted in the main text, we have also verified that we can measure non-zero
values for \ion{Ca}{14} 193.874\,\AA\, in an X1.8 solar flare when the line is more prominent
above the background. This supports the arguments made here.

\bibliography{ms}
\bibliographystyle{ms}

\end{document}